\documentclass{emulateapj}
\usepackage{epsfig} 
\usepackage{natbib}
\usepackage{epstopdf}
\usepackage{graphicx}
\usepackage{epstopdf}

\shortauthors{Bochanski et al.} \shorttitle{The Most Distant Stars in the Galaxy} 
\begin{document}

\title{Hunting The Most Distant Stars in the Milky Way:  Methods and Initial Results}

\author{ John J. Bochanski\altaffilmark{1}, Beth Willman\altaffilmark{1}, Andrew A. West\altaffilmark{2}, Jay Strader\altaffilmark{3}, Laura Chomiuk\altaffilmark{3}} 

\altaffiltext{1}{Haverford College, 370 Lancaster Ave, Haverford PA 19041 USA\\
email:jbochans@haverford.edu} 
\altaffiltext{2}{Department of Astronomy, Boston University, 725 Commonwealth Avenue, Boston, MA 02215 USA} 
\altaffiltext{3}{Michigan State Astronomy Group, Michigan State University, Biomedical Physical Sciences Building, 567 Wilson Road, Room 3261
              East Lansing, MI  48824-2320 USA}

\begin{abstract}
We present a new catalog of 404 M giant candidates found in the UKIRT Infrared Deep Sky Survey (UKIDSS).   The 2,400 deg$^2$ available in the UKIDSS Large Area Survey Data Release 8 resolve M giants through a volume four times larger than that of the entire Two Micron All Sky Survey.  Combining near-infrared photometry with optical photometry and proper motions from the Sloan Digital Sky Survey yields an M giant candidate catalog with less M dwarf and quasar contamination than previous searches for similarly distant M giants.  Extensive follow-up spectroscopy of this sample will yield the first map of our Galaxy's outermost reaches over a large area of sky.  Our initial spectroscopic follow-up of $\sim$ 30 bright candidates yielded the positive identification of five M giants at distances $\sim 20-90$ kpc. Each of these confirmed M giants have positions and velocities consistent with the Sagittarius stream. The fainter M giant candidates in our sample have estimated photometric distances $\sim 200$ kpc (assuming $[Fe/H]$ = 0.0), but require further spectroscopic verification.  The photometric distance estimates extend beyond the Milky Way's virial radius, and increase by $\sim 50\%$ for each 0.5 dex decrease in assumed $[Fe/H]$. Given the number of M giant candidates, initial selection efficiency, and volume surveyed, we loosely estimate that at least one additional Sagittarius-like accretion event could have contributed to the hierarchical build-up of the Milky Way's outer halo.

\end{abstract}

\section{Introduction} 

Our current understanding of the formation and evolution of the Milky Way (MW) is intimately tied to the $\Lambda$-Cold Dark Matter (CDM) paradigm:  large galaxies such as our own were assembled from smaller galaxies over a Hubble time \citep[e.g.,][]{1978ApJ...225..357S}.  This accretion process is most directly observable in the Galactic halo, where dynamical times are long and structures can be kinematically and spatially observed at the present date \citep{2005ApJ...635..931B,2008ApJ...689..936J}.  This theoretical model of hierarchical formation has strong empirical support from the numerous, spatially coherent tidal streams mapped in studies based on Sloan Digital Sky Survey \citep[SDSS;][]{2000AJ....120.1579Y} and the Two--Micron All Sky Survey \citep[2MASS;][]{2006AJ....131.1163S} observations \citep{2002ApJ...569..245N,2003ApJ...588..824Y,2004A&A...423..517R,2006MNRAS.365.1309H,2006ApJ...642L.137B,2007ApJ...654..897B,2009ApJ...693.1118G,2010ApJ...722..750S,2012ApJ...760L...6B,2013ApJ...769L..23G}.  In addition to providing a direct measure of our Galaxy's formation history, observations of RR Lyrae and blue horizontal branch stars in the MW's halo have been used to constrain both the amount \citep[e.g.,][]{wilkinson99a, 2008ApJ...684.1143X, 2012MNRAS.425.2840D, gnedin10a} and the distribution of the Milky Way's total mass \citep{deason12a}.

Simulations show that the MW's outer halo is particularly interesting for near field cosmology, because the greatest stellar differences between MW models with different merging histories are predicted to be at $d \gtrsim$ 40 kpc \citep{2008ApJ...689..936J,2010MNRAS.406..744C}.  For example, observations of a sharp break in the MW halo's density profile at 20 $< d <$ 30 kpc \citep[e.g.,][]{2009MNRAS.398.1757W,2010ApJ...708..717S,2013ApJ...766...24D}, may suggest that the Milky Way's halo formation was dominated by a single relatively massive satellite \citep{deason13a}.  Dynamical times are also very long in the outer halo \citep[$> 15$ Gyr,][]{2008gady.book.....B}, resulting in the predicted dominance of spatially correlated stellar streams from dwarf merging events.  Identification of these streams to the outermost reaches of the stellar halo is necessary to include the most eccentric, and also the most recent, merger events in our picture of the MW's formation history \citep{sharma11a}.

Unfortunately, we currently have a largely incomplete view of the outer stellar halo.  Its outermost reaches ($d \gtrsim 150$ kpc) have not yet been mapped over a wide area of sky.  While old, metal-poor main sequence turnoff stars have revolutionized our understanding of the MW's inner halo \citep[e.g.,][]{bell08a}, they cannot yet be observed at $d \gtrsim$50 kpc over large areas of sky.  More luminous tracers, such as red giant branch stars \citep[e.g.,][]{helmi03a,xue12a}, RR Lyrae \citep[e.g.,][]{sesar12a,drake13a}, and blue horizontal branch stars \citep[BHBs, e.g.][]{yanny00a,schlaufman09a} have been used to map the Galactic halo to distances out to $d \sim 120$ kpc.  One study selected M giant candidates at significant distances $d > 100$ kpc from the SDSS photometric catalog \citep{2012AJ....143..128P}, however the low photometric S/N ($<$ 5) and the lack of spectroscopic follow-up of these candidates significantly limits the interpretive value of that candidate sample.
  
M giants provide a mechanism to map the Galaxy's halo to its outermost limits over a relatively wide field.  They are extremely bright, with typical luminosities of $1 \times 10^3 L_{\odot}$.  Unlike earlier type giant stars, M giants can be photometrically separated from foreground dwarf stars using a combination of near--infrared (NIR) and optical colors \citep{1970ApJ...162..217L,1975MNRAS.171P..19G,1976ApJ...210..402M, 1988PASP..100.1134B}.  The ongoing UKIDSS Large Area Survey (LAS, \citealp{2007MNRAS.379.1599L}) provides the opportunity to map M giants beyond the Milky Way's virial radius ($d\gtrsim$200 kpc) over nearly $1/20$ of the sky.  Explained in detail below, the UKIDSS Large Area Survey Data Release 8 (LAS DR8) covers 2,400 deg$^2$, will ultimately cover 4,000 deg$^2$, and extends 3-4 magnitudes fainter than 2MASS.  Despite the smaller surface area coverage, UKIDSS LAS DR8 alone probes a volume of the halo four times greater than the entirety of 2MASS.  

M giants are a relatively biased tracer of our halo, preferentially revealing the past merging of the relatively high-luminosity ($L \gtrsim 5 \times 10^6 L_{\odot}$) accretion events predicted to contribute the most to the stellar halo's luminosity \citep{bullock05a,sharma11a}.  Stars with $[Fe/H] \gtrsim -0.8$ and intermediate ages ($\gtrsim 5$ Gyr) have low effective temperatures \citep[$< 4000$ K;][]{1999AJ....117..521V} and are classified as M giants \citep{2008ApJS..178...89D}.  Past observational work has amply demonstrated that M giants are effective tracers of relatively high luminosity accretion events.  M giants selected using 2MASS NIR \citep{2003ApJ...599.1082M} and SDSS optical colors \citep{2009ApJ...700.1282Y} have been used to trace out the Sgr dwarf's tidal debris across the entire sky.  Other NIR selected M giants revealed the presence of a possible past accretion event in Canis Major \citep{martin04a}, and have been used to search for loose stellar associations in the Milky Way's halo, out to distances of $d < 50$ kpc \citep{ibata02a,2010ApJ...722..750S}.  

In this paper, we present our set of outer halo star candidates, with our sample distances extending beyond the virial radius of the Milky Way over $>1/20$ of the sky.  We identify and characterize M giants using a combination of UKIDSS NIR and SDSS optical colors, along with proper motions.  In Section \ref{sec:obs}, we summarize the UKIDSS and SDSS photometric and spectroscopic catalogs.  Section \ref{sec:targ_selection} describes the selection process of our M giant candidates.  Since M giants are susceptible to contamination from background quasars and foreground M dwarfs, we describe the results our our initial spectroscopic campaign in Section \ref{sec:results}.  We characterize the spatial distribution of M giant candidates on the sky, along with distance estimates in Section \ref{sec:discussion}. We discuss implications for the progenitors of these stars, paths for future investigations and conclusions in Section \ref{sec:conclusions}.

\section{Survey Observations}\label{sec:obs}
The survey observations used in our investigation are described below.  M giant candidates were selected using NIR colors, and matched SDSS optical photometry and astrometry from SDSS/USNO-B was employed to remove quasars and M dwarfs from our final sample.

\subsection{UKIDSS}
UKIDSS \citep{2007MNRAS.379.1599L} is a NIR photometric survey being conducted at the 3.8m UK Infrared Telescope (UKIRT). The survey employs the UKIRT Wide Field Camera \citep[][]{2007A&A...467..777C}, imaging the sky with a $zyJHK$ filter set.
The UKIDSS survey is comprised of 5 smaller surveys, each with specific science goals. The largest component is the LAS which will ultimately include $\sim$ 4,000 deg$^2$ to a faint limit of $K \sim 18.2$, nearly four magnitudes deeper than the 2MASS completeness limit.  In \S\ref{sec:nircolor}, we motivate limiting this sample to sources with a signal--to--noise (S/N) ratio of $>$ 15, where the survey reaches depths of 18.47,17.66,17.05 in $J,H,K$ (Vega mag), respectively.   The latest public release\footnote{The UKIDSS data is available at \url{http://http://surveys.roe.ac.uk/wsa/}.} (DR8) of the LAS contains $\sim 2,400$ sq. deg. of $yJHK$ imaging, using a typical exposure time of 40s.  

\begin{figure*} 
  \centering 
\includegraphics[scale=0.4]{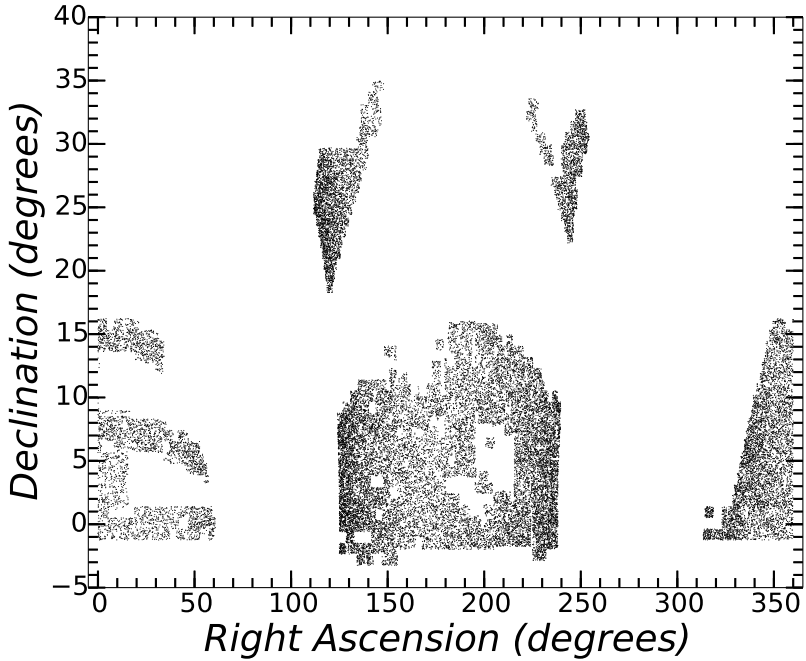}
  \caption{The footprint of the UKIDSS LAS publicly available in DR8 \citep{2012yCat.2314....0U} in right ascension and declination. UKIDSS LAS DR8 spans 2,400 deg$^2$ and contains over 69 million sources.}
  \label{fig:survey_geo} 
 \end{figure*}

The astrometric precision for UKIDSS imaging varies with brightness and Galactic latitude, but is better than $\sim 0.1$ arc second in each coordinate \citep{2007MNRAS.379.1599L,2009MNRAS.394..857D,2010A&A...515A..92S} over the majority of the survey. UKIDSS astrometry, combined with 2MASS and SDSS positions, have already yielded proper motions for nearby stars \citep{2010A&A...515A..92S}. The DR8 LAS sky coverage is shown in Figure \ref{fig:survey_geo}.

We corrected the $JHK$ magnitudes of all sources for reddening using the extinction maps of \cite{1998ApJ...500..525S}, with the updated coefficients from \cite{2011ApJ...737..103S}.   The reddening in our final sample was small, with a median $E(B-V) = 0.048$ and a standard deviation of $\sigma_{E(B-V)} = 0.043$.  The maximum reddening in the sample was $E(B-V) = 0.41$, below the limit of $E(B-V) = 0.55$ set by \cite{2003ApJ...599.1082M}.  

\subsection{Sloan Digital Sky Survey} 
The SDSS \citep[][]{2000AJ....120.1579Y} was an optical photometric and spectroscopic survey of 14,555 deg.$^2$ \citep{2013arXiv1307.7735A} conducted with the 2.5m telescope at the Apache Point Observatory \citep{2006AJ....131.2332G}. The bulk of the survey was photometric, surveying the sky using a drift-scan technique. The camera contained 5 camera columns, each corresponding to one filter in the $ugriz$ system. With an exposure time of 53.9 seconds, the main survey achieved a faint limit of $r \sim$ 22, with typical systematic photometric errors of $\sim$ 0.02 mag for point sources \citep{2007AJ....134..973I}. This was empirically verified using repeat scans of a 300 sq. deg. area near $\delta = 0$, the so-called ``Stripe 82'', which was imaged multiple times over the course of the survey \citep[i.e.,][]{2011ApJ...731...17B}.

When the skies above Apache Point were not photometric, the SDSS operated in spectroscopic mode. The spectrograph contained two cameras which simultaneously obtained $R \sim 1,800$ spectra for 640 objects through fiber optic cables \citep{1999AAS...195.8701U}. The typical exposure time was $\sim$ 45 minutes, resulting in many observations with S/N $> 15$.  As part of the Legacy and SEGUE surveys, stars were selected for spectroscopic followup by targeting algorithms designed to retrieve specific types of stars, such as cataclysmic variables \citep{2002AJ....123..430S}, brown dwarfs \citep{2010AJ....139.1808S},  K giants \citep{helmi03a} and a host of $\sim$15 other types of stars \citep{2009AJ....137.4377Y}.  Over 3 million unique optical spectra are publicly available in the latest data release \citep{2013arXiv1307.7735A}, including more than 600,000 stellar spectra.  Recently, the Apache Point Observatory Galactic Evolution Experiment \citep[APOGEE,][]{2013AJ....146...81Z} obtained and published $\sim$ 60,000 NIR spectra of red giant stars, although these stars are much brighter than the ones presented here ($H \lesssim 12$).

The SDSS also delivered exquisite astrometry, with an internal precision of 25 mas and an absolute accuracy of 45 mas in each coordinate direction \citep{2003AJ....125.1559P}. A proper motion catalog was derived by comparing SDSS observations to the USNO-B catalog \citep{2003AJ....125..984M}, with a typical baseline of 50 years. The resulting catalog \citep{2004AJ....127.3034M} contains over 205 million objects with a median uncertainty of 2 mas yr$^{-1}$. These proper motions have been used to study a number of stellar populations, including nearby M dwarfs and subdwarfs \citep{2007AJ....134.2418B,2013AJ....145...40B, 2008ApJ...681L..33L}.

\section{M Giant Candidate Selection}\label{sec:targ_selection}
To identify the stars most likely to be M giants, we incorporate UKIDSS and SDSS photometry into our targeting algorithm.  We also use proper motion estimates to cull out nearby M dwarfs.  The process is explained below and summarized in Table \ref{table:target_cuts}.  

\subsection{Near-Infrared Color Cuts}\label{sec:nircolor}
The UKIDSS LAS DR8 catalog contains 69,656,410 sources, down to a faint limit of $K \sim 18.2$, covering $\sim 2,400$ sq. deg.  We implemented a SQL query on the UKIDSS public database designed to only include high-quality detections of point sources (at the expense of completeness), yielding 6,969,163 objects.  Our SQL query is available in Appendix A.   From the catalog of point sources, we selected M giant candidates using the following NIR color-color cuts:
\begin{equation}\label{eqn:colorcut1}
	(J-K)_{\rm o}	 > 1.02	
	\end{equation}
\begin{equation}
		(J-H)_{\rm o}	 < 0.561 \times (J-K)_{\rm o} + 0.46	
\end{equation}
\begin{equation}
		(J-H)_{\rm o}	 > 0.561 \times (J-K)_{\rm o} + 0.185	
\end{equation}

This color selection was adapted from those of \cite{2003ApJ...599.1082M} and \cite{2010ApJ...722..750S}, but shifted slightly redward (in both $J-H$ and $J-K$), to reduce contamination from foreground M dwarfs.  The most significant difference between our own selection and previous studies is contained in Eqn. \ref{eqn:colorcut1}.  The M dwarf and M giant sequence begin to diverge in $JHK$ colors near $J-K \sim$ 0.85, with larger separation at redder colors \citep{1970ApJ...162..217L,1975MNRAS.171P..19G,1976ApJ...210..402M, 1988PASP..100.1134B}.    By extending our color cuts redward, we maximize the separation from the locus of M dwarfs, minimizing the contamination of our M giant sample.   Our NIR color selection is shown in Figure \ref{fig:nir_color}. This color selection is most sensitive to M giants with $[Fe/H] > -0.5$ \citep{2008ApJS..178...89D}.

Since the number of M dwarfs is much larger than the number of M giants in the Galaxy, we estimated our contamination rate by comparing the number of M dwarfs expected to scatter into our selection box as a function of signal--to--noise (S/N) and color.  In order to facilitate this computation, we computed the expected scatter as a function of $c_1$ color, which is a linear combination of $J-H, H-K$ colors defined as:
\begin{equation}\label{eqn:c1}
c_1 = (J-H)-(0.561*(J-K))
\end{equation}
This color effectively collapses the NIR colors along the slope of NIR giant selection box shown in Figure \ref{fig:nir_color}.  M giants candidates have red $c_1$ colors, while M dwarfs and brown dwarfs have bluer $c_1$ colors.  In steps of S/N = 5 from 5 to 100, we fit the low-mass stellar locus with a Gaussian, then computed the expected number of low--mass stars scattered into the M giant NIR selection box.  
The contamination level dropped quickly with rising S/N as: 1) The number of M dwarfs decreased and; 2) the separation between the M dwarf and M giant loci increased.  
We also calculated the expected contamination while adopting the \cite{2003ApJ...599.1082M} and \cite{2010ApJ...722..750S} NIR color cuts.  These previous studies were subject to greater M dwarf contamination due to their bluer $JHK$ color cuts.     Our adopted cuts emphasize purity over completeness, avoiding the bluer $JHK$ color regions which do include M giants \citep{2003ApJ...599.1082M,2004AJ....128..245M} but are dominated by M dwarf contamination.  After this experimentation, we implemented a S/N $=$ 15 for imaging in $JHK$, which translates to faint limits of $J,H,K = (18.47,17.66,17.05)$.  At this fiducial S/N, we expect $\sim 25\%$ contamination from M dwarfs. However, the fraction of M dwarfs contaminating our final sample is larger, as detailed in Section \ref{sec:contam}.   This cut resulted in 1,649 M giant candidates with NIR photometry.  

\subsection{Optical Color Cuts}\label{sec:optical}
Using the SDSS \textsc{casjobs} interface \citep{2005cs........2072O}, we cross-matched our list of M giant candidates against the SDSS DR10 \textsc{PhotoPrimary} catalog\footnote{The DR9 casjobs website is available at \url{http://skyserver.sdss3.org/CasJobs/}.}.  Of the 1,649 NIR selected objects, 1,638 matched objects in SDSS within 30$^{\prime\prime}$.  Although we permitted a large matching radius, the majority of stars (99.1\%) had matches within 1$^{\prime\prime}$, with only 9 sources outside this radius.  All of these distant matches were removed with proper motion cuts.  We obtained $ugriz$ psf photometry, flags, and uncertainties for each object, as well as any available proper motions and spectroscopy.  Of the matched UKIDSS-SDSS observations that passed our visual vetting (see Section \ref{sec:visualvetting}), 98 were removed for having bad photometry in SDSS (\textsc{clean} = 0) or being morphologically identified as a galaxy (\textsc{type} = 3). 

\subsection{Visual vetting}\label{sec:visualvetting}
During the course of our investigation, we learned of issues present in a fraction of the stacked $J$ band frames that result in  $J$ magnitudes that are systematically faint by $\sim$0.14 mag (S. Warren, private communication, 2012). Unfortunately, no flags in the catalog alert the user to this anomaly.  Such dimmed $J$ magnitudes make the $J-H$ and $J-K$ colors of affected stars artificially redder.  Stars on flawed frames can be tossed out via visual inspection: the artifact is manifested as pixel-sized ``holes'' in the $J$ band imaging, equally spaced by a few pixels.  These systematic errors affect entire $J$-band frames, with every star on a bad frame being dimmed.  An example of a good and bad $J$ band image is shown in Figure \ref{fig:cutout}.  To achieve the highest fidelity, we visually inspected each matched SDSS-UKIDSS M giant candidate.  Of the 1,638 matches to SDSS photometry, 992 were obtained with clean $J$-band frames.  The remaining candidates were removed from our sample.
\begin{center}
\begin{deluxetable*}{lcr}
\tablewidth{0pt}
\tabletypesize{\small}
 \tablecaption{Summary of Target Selection}
 \tablehead{
 \colhead{Dataset} &
 \colhead{Description} &
 \colhead{Stars Remaining} 
}
 \startdata
UKIDSS & Total UKIDSS DR8 LAS Footprint & 69,656,410 \\
UKIDSS & High quality imaging & 6,969,163\\
UKIDSS & M Giant Color-Cuts, S/N = 15 & 1,649 \\
UKIDSS $+$ SDSS & NIR $+$ Optical Match & 1,638 \\
UKIDSS $+$ SDSS  & Visual Inspection & 992 \\
UKIDSS $+$ SDSS  & Optical Color Cuts & 464 \\
UKIDSS $+$ SDSS  & USNO-B-SDSS Proper Motions & 409 \\
UKIDSS $+$ SDSS  & SDSS-UKIDSS Proper Motions & 404 
\enddata
 \label{table:target_cuts}
\end{deluxetable*}
\end{center}

 \begin{figure} 
  \centering 
 \includegraphics[scale=0.3]{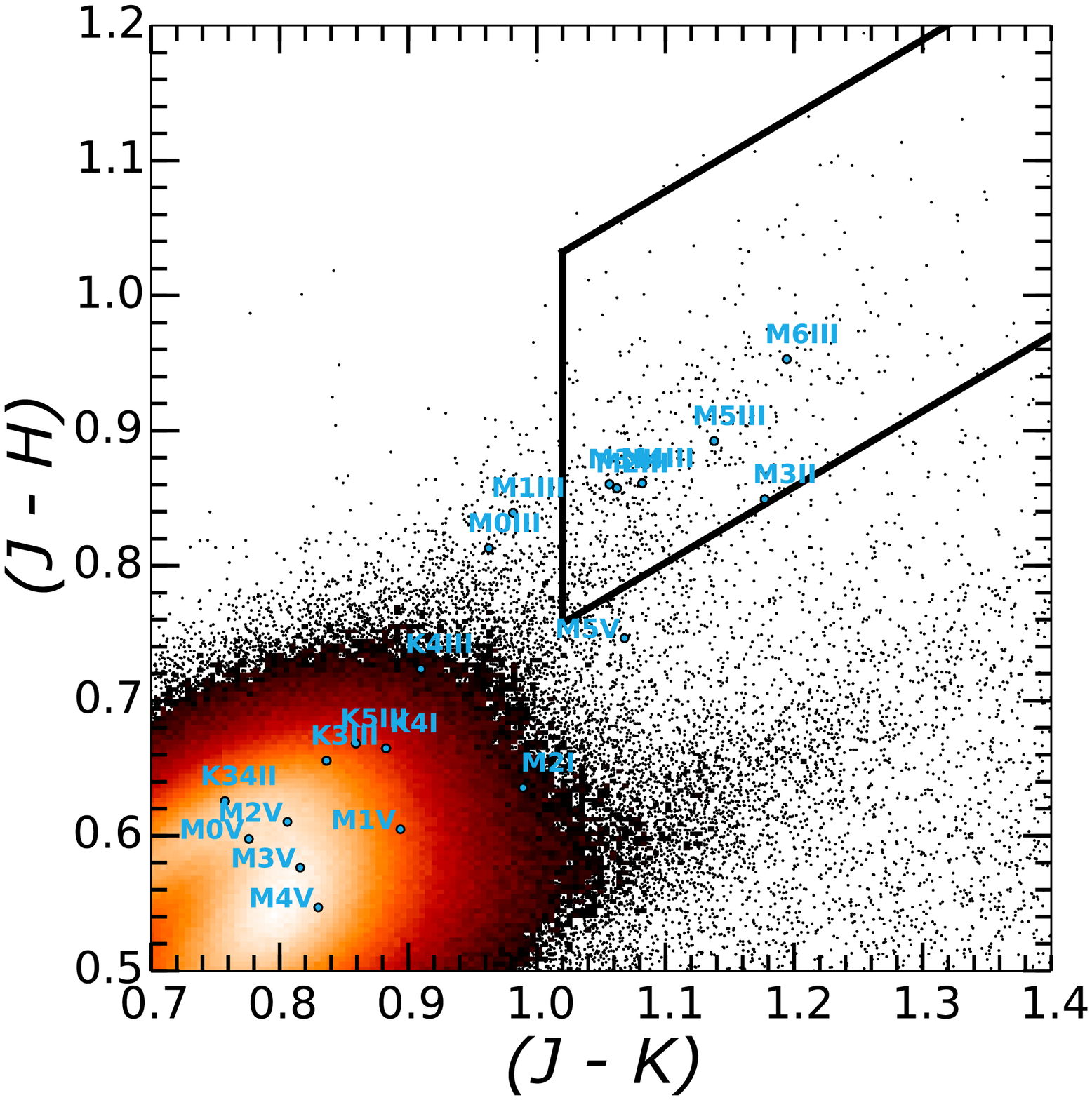}

 \caption{The $J-H, J-K$ (Vega) color-color diagram for point sources in UKIDSS.  Overplotted is our NIR color selection described in Equation \ref{eqn:colorcut1} and the colors of M giants and dwarfs synthesized from the Pickles spectral library \citep[light blue points,][]{1998PASP..110..863P}.  Our NIR color selection is shifted to slightly redder $J-H, J-K$ colors than that of \cite{2003ApJ...599.1082M} and \cite{2010ApJ...722..750S}.  This was done to reduce contamination of bluer, nearby M dwarfs, which dominate the stellar locus at bluer colors.  This may reject some early--type giants, but should significantly reduce dwarf contamination.}
  \label{fig:nir_color} 
 \end{figure}

 \begin{figure*} 
  \centering 
\includegraphics[scale=0.4]{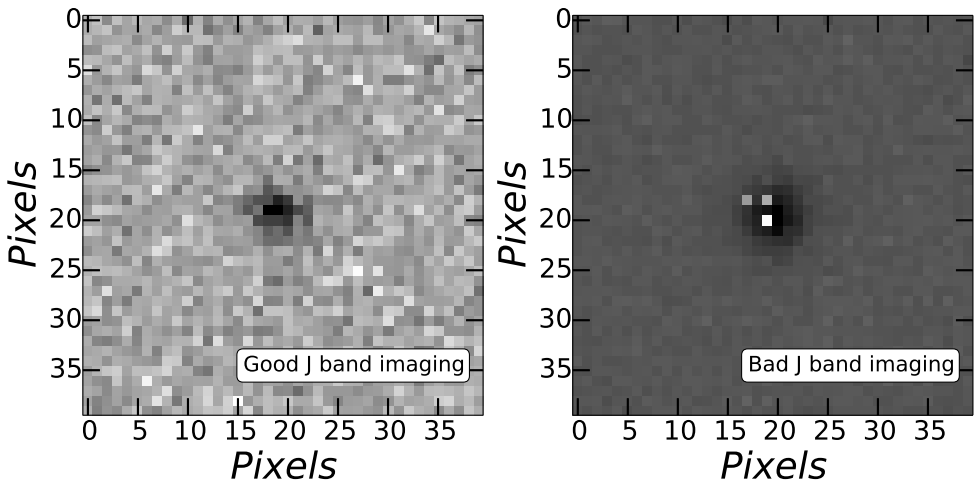}

  \caption{An example of ``bad'' imaging in UKIDSS.  The cutout on the left is a star that passed our visual verification, while the star on the right contained faulty $J$--band imaging and was rejected.  Images were readily identified by ``holes'' offset by a few pixels, typically near bright sources.  This effect removed 646 M giant candidates from our study.}
  \label{fig:cutout} 
 \end{figure*}
 
\subsection{Removing QSO Contamination}
\begin{figure} 
  \centering 
\includegraphics[scale=0.3]{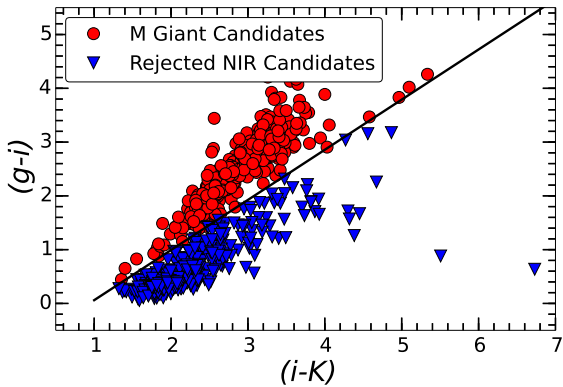}
 
 \caption{The NIR-optical $g-i,i-K$ color-color selection used to remove QSOs (blue filled triangles) from our M giant candidates (red filled circles).  Note that the QSOs have bluer $g-i$ colors at the same $i-K$.
     The combination of optical and NIR colors are essential for
     removing QSOs from the sample, which may have caused minor contamination in previous studies \citep[i.e.,][]{2003ApJ...599.1082M,2010ApJ...722..750S}.}
  \label{fig:optical_cut} 
 \end{figure}

While the surface density of QSOs is small, they have NIR colors similar to M giants.  Fortunately, NIR and optical colors can discriminate between the two classes.  \cite{2011AJ....141..105P} demonstrated that stars and QSOs separate in $g-i, i-K$ color-color space, as shown in Figure \ref{fig:optical_cut}.  We require our M giant candidates to satisfy:

\begin{equation}
g-i > 0.932 \times (i-K) - 0.872
\end{equation}\label{eqn:qso}

In the current sample, this cut removes nearly half of NIR selected candidates, with 466 M giant candidates remaining (and 428 removed).  Finally, we enforced red optical colors, by requiring $g-r > 0.5$ for all candidates, removing two objects from the catalog.

\begin{figure} 
  \centering 
\includegraphics[scale=0.3]{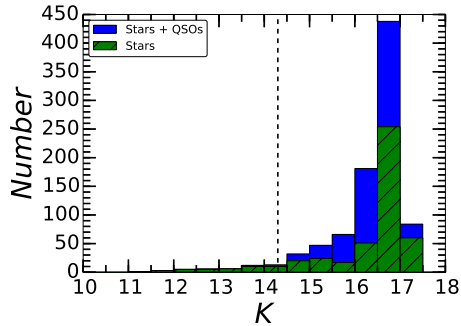}
 
 \caption{The $K$ magnitude distributions of stars (hatched green histogram) and all objects (QSOS and stars; solid purple histogram) in our sample.  Note that QSOs become more important at fainter magnitudes, well beyond the faint limit of 2MASS ($K=14.3$).  This suggests that QSO contamination was not significant for previous studies that relied solely on photometric identification of M giants \citep[i.e.,][]{2003ApJ...599.1082M, 2010ApJ...722..750S}.}
  \label{fig:qso_hist} 
 \end{figure}

Since nearly half of our sample is removed with this cut, we initially suspected QSO contamination may have affected previous M giant studies. We investigated the effect of QSO contamination on catalogs of 2MASS--selected M giants \citep[i.e.,][]{2003ApJ...599.1082M, 2010ApJ...722..750S}.   In Figure \ref{fig:qso_hist}, we plot the apparent $K$ distribution of stars and stars and quasars, as defined by Equation \ref{eqn:qso}.  As seen in Figure \ref{fig:qso_hist}, QSO contamination becomes important after $K \sim 14.5$, which is just beyond the faint limit of 2MASS ($K = 14.3$).  Although there are few UKIDSS sources bright enough to robustly test the possible impact of QSO contamination on 2MASS studies, our sample implies that previous studies would not have been significantly influenced by the presence of QSOs.

Our cut to remove QSOs was tested during an initial observing run for this project.  We obtained Gemini GNIRS spectra (described in Section \ref{sec:gemini}) of 14 M giant candidates selected with an old target selection algorithm that did not include proper motion or optical colors.  Of the 14 targets, eight had $giK$ colors that would have been flagged as QSOs by Eqn. \ref{eqn:qso}.  All eight were spectroscopically confirmed as $z \sim 1.2$ QSOs, suggesting this cut is very effective at removing these contaminants.

\subsection{Proper Motion Cuts}
M giants at halo distances should exhibit no discernible proper motion.  Thus, we removed any stars with significant proper motions using the two following methods.  The first used the proper motions measured between the SDSS and USNO-B catalog \citep{2004AJ....127.3034M,2003AJ....125..984M}, which was based on observations mainly from the POSS survey \citep{1963bad..book..481M}.  Proper motions, when available, were obtained from the \textsc{ProperMotions} table
in SDSS DR10.  With a typical baseline of 50 years, the precision of proper motion estimates was $\sim 2$ mas yr$^{-1}$.
We used several cuts on precision
and error flags to ensure high quality proper motions.
These are described in detail in our previous investigations
\citep[i.e.,][]{2010AJ....139.2566D, 2011AJ....141...98B, 2011AJ....141...97W}.  
Any star that exhibited a proper motion larger than 2.5$\times$ the uncertainty in either right ascension or declination was flagged as moving.  This cut removed 55 stars from our sample.

\begin{figure} 
  \centering
\includegraphics[scale=0.4]{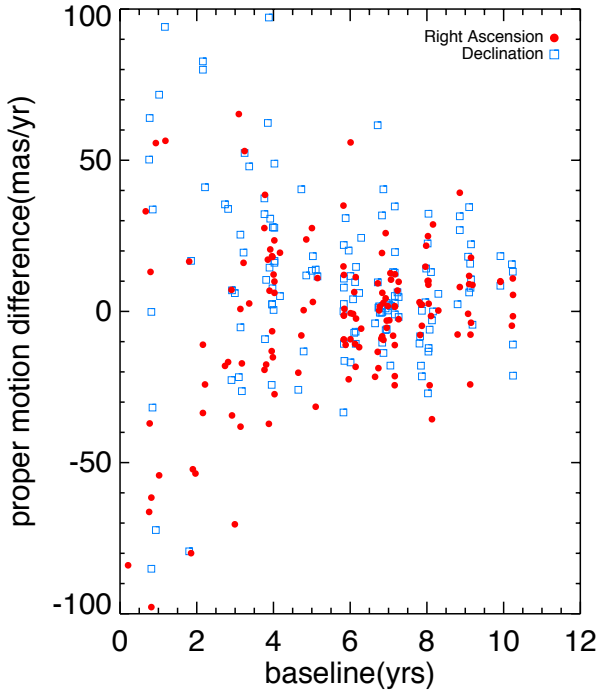}
 
 \caption{The difference in proper motion as measured by \cite{2004AJ....127.3034M} and our SDSS-UKIDSS proper motions as a function of SDSS-UKIDSS temporal baseline.  Proper motion differences in right ascension (filled red circles) and declination (open blue squares) follow similar distributions, with better agreement for longer baselines.  The standard deviation in proper motion difference for baselines greater than six years is $\sim$ 40 mas yr$^{-1}$, which we adopt as our precision limit for SDSS-UKIDSS proper motions.  Stars with motions greater than 40 mas yr$^{-1}$ in either direction, and with baselines greater than 6 years, were expunged from the sample.  This removed five stars from our sample.}
  \label{fig:baseline}  
 \end{figure}

Since the \cite{2004AJ....127.3034M} proper motions are based on POSS observations, they lack sensitivity to faint red objects.  We addressed this by computing proper motions using SDSS as the first epoch and UKIDSS as the second epoch.
The astrometric positions are reported by the
SDSS pipeline with an internal precision of 25 mas and an absolute
accuracy of 45 mas in each coordinate direction \citep{2003AJ....125.1559P}.  The UKIDSS astrometric precision ranges from 50-100 mas, depending on Galactic latitude \citep{2007MNRAS.379.1599L,2010A&A...515A..92S}.  We computed proper motions for each star using the UKIDSS and SDSS positions, and compared them to the \cite{2004AJ....127.3034M} proper motions as a function of temporal baseline (Figure \ref{fig:baseline}).  After $\sim 6$ years, the proper motions agree at a level of $\sim 40 $ mas yr$^{-1}$.  Any stars with a motion greater than 40 mas yr$^{-1}$ (in either coordinate) with a baseline of 6 years or greater were excluded from our sample, with 5 stars being removed.

\begin{figure} 
  \centering 
\includegraphics[scale=0.3]{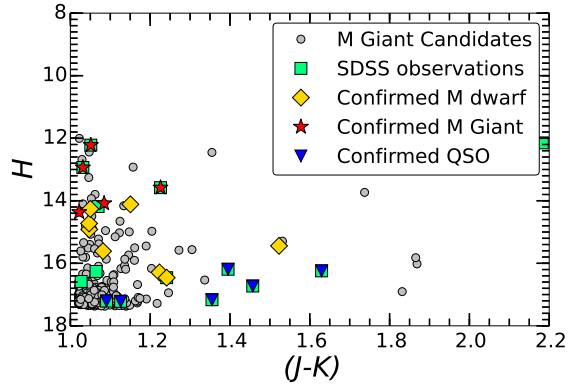}
  
\caption{The $J-K, H$ color--magnitude diagram for our final M giant candidate sample.  Spectroscopically confirmed giants are shown with red filled stars,  M dwarfs are shown with yellow filled diamonds, and spectroscopically identified QSOs are shown with blue filled triangles.  Most of our candidates have 1.0 $< (J-K)_o <$ 1.2, and our confirmed M giants have $H$ brighter than 15.  We expect more distant M giants will be found as we followup these targets with larger telescopes.}
  \label{fig:cand_cmd} 
 \end{figure}

The GNIRS spectra were also useful for testing the effectiveness of our proper motion cuts.  Of the six stars observed with GNIRS, two were flagged as moving, while four were not.  All were spectroscopically confirmed as low--mass dwarfs, suggesting an efficiency of $\sim 30\%$.  After the NIR, optical and proper motion cuts, 404 stars remained in our sample.  These stars are listed in Table \ref{table:candidates} and their NIR color--magnitude diagram is shown in Figure \ref{fig:cand_cmd}.  Our initial spectroscopic followup is detailed in the following section.

\begin{turnpage}
\begin{center}
\begin{deluxetable*}{lllllllllllllllllll}
\tablewidth{0pt}
\tabletypesize{\scriptsize}
 \tablecaption{M Giant Candidates}
 \tablehead{
 \colhead{Name} &
 \colhead{RA} &
 \colhead{Dec} &
 \colhead{$u$} &
\colhead{$\sigma_u$} &
 \colhead{$g$} &
\colhead{$\sigma_g$} &
 \colhead{$r$} &
\colhead{$\sigma_r$} &
 \colhead{$i$} &
\colhead{$\sigma_i$} &
 \colhead{$z$} &
\colhead{$\sigma_z$} &
 \colhead{$J$} &
\colhead{$\sigma_J$} &
 \colhead{$H$} &
\colhead{$\sigma_H$} &
 \colhead{$K$}  &
\colhead{$\sigma_K$} 
}
 \startdata
ULAS J000216.74+115533.0  &   0.56976  &  11.92584 & 25.32 &  0.67 & 24.93 &  0.49 & 22.36 &  0.16 & 20.33 &  0.04 & 19.17 &  0.06 & 17.46 &  0.03 & 16.73 &  0.03 & 16.44 &  0.04 \\
ULAS J001228.63+003049.8  &   3.11930  &   0.51384 & 23.52 &  0.79 & 21.77 &  0.06 & 20.31 &  0.03 & 19.03 &  0.02 & 18.30 &  0.03 & 17.16 &  0.02 & 16.40 &  0.03 & 16.10 &  0.03 \\
ULAS J001507.92+070058.8  &   3.78301  &   7.01634 & 25.37 &  0.76 & 23.05 &  0.22 & 21.58 &  0.08 & 19.97 &  0.03 & 19.04 &  0.05 & 17.70 &  0.05 & 16.90 &  0.05 & 16.66 &  0.06 \\
ULAS J001535.72+015549.6  &   3.89882  &   1.93045 & 25.03 &  1.02 & 21.16 &  0.04 & 19.80 &  0.03 & 19.12 &  0.02 & 18.74 &  0.04 & 17.73 &  0.06 & 17.00 &  0.04 & 16.70 &  0.05 \\
ULAS J001701.41+081837.4  &   4.25585  &   8.31038 & 22.64 &  0.59 & 22.59 &  0.22 & 21.73 &  0.12 & 20.11 &  0.04 & 19.09 &  0.07 & 17.74 &  0.04 & 16.98 &  0.05 & 16.69 &  0.05 \\
ULAS J001827.65+142946.6  &   4.61523  &  14.49629 & 22.73 &  0.56 & 20.02 &  0.03 & 18.69 &  0.02 & 17.93 &  0.02 & 17.45 &  0.02 & 16.49 &  0.01 & 15.67 &  0.02 & 15.43 &  0.02 \\
ULAS J002219.76+073542.6  &   5.58232  &   7.59516 & 23.09 &  0.38 & 21.80 &  0.06 & 20.55 &  0.03 & 19.62 &  0.02 & 19.15 &  0.04 & 18.11 &  0.08 & 17.34 &  0.06 & 17.02 &  0.07 \\
ULAS J002226.36+071711.8  &   5.60981  &   7.28662 & 24.39 &  0.63 & 23.49 &  0.21 & 21.94 &  0.07 & 20.24 &  0.03 & 19.41 &  0.05 & 17.89 &  0.05 & 17.16 &  0.05 & 16.87 &  0.06 \\
ULAS J002548.91+150745.0  &   6.45381  &  15.12918 & 23.71 &  0.75 & 23.41 &  0.36 & 21.37 &  1.63 & 19.95 &  0.77 & 19.21 &  0.19 & 17.74 &  0.06 & 16.87 &  0.05 & 16.65 &  0.07 \\
ULAS J003048.68-005859.0  &   7.70284  &  -0.98305 & 23.15 &  0.45 & 20.56 &  0.03 & 19.33 &  0.02 & 18.77 &  0.02 & 18.50 &  0.03 & 17.50 &  0.05 & 16.72 &  0.04 & 16.46 &  0.06 \\
ULAS J003105.14+144525.5  &   7.77140  &  14.75709 & 24.86 &  0.74 & 22.89 &  0.17 & 21.48 &  0.07 & 20.07 &  0.03 & 19.26 &  0.06 & 17.95 &  0.06 & 17.15 &  0.06 & 16.93 &  0.08 \\
ULAS J003448.36+142924.6  &   8.70150  &  14.49016 & 23.29 &  0.56 & 21.93 &  0.08 & 20.52 &  0.04 & 19.65 &  0.03 & 19.16 &  0.05 & 18.12 &  0.07 & 17.31 &  0.08 & 17.01 &  0.08 \\
ULAS J003501.38+150324.4  &   8.75574  &  15.05677 & 23.82 &  0.90 & 20.87 &  0.04 & 19.53 &  0.02 & 18.82 &  0.02 & 18.47 &  0.04 & 17.40 &  0.03 & 16.60 &  0.04 & 16.38 &  0.05 \\
ULAS J004216.12+084509.9  &  10.56717  &   8.75276 & 22.95 &  0.69 & 23.91 &  0.74 & 21.94 &  0.13 & 20.32 &  0.05 & 19.45 &  0.09 & 17.95 &  0.04 & 17.17 &  0.06 & 16.91 &  0.07 \\
ULAS J004255.06+065205.1  &  10.72943  &   6.86808 & 23.29 &  0.75 & 21.44 &  0.06 & 20.02 &  0.03 & 19.36 &  0.03 & 18.86 &  0.05 & 17.91 &  0.05 & 17.11 &  0.05 & 16.85 &  0.07 \\
ULAS J004505.03+141010.8  &  11.27094  &  14.16967 & 22.65 &  0.44 & 22.61 &  0.19 & 21.35 &  0.07 & 20.01 &  0.03 & 19.30 &  0.06 & 18.01 &  0.05 & 17.26 &  0.06 & 16.94 &  0.07 \\
ULAS J004823.61+135332.6  &  12.09837  &  13.89238 & 25.42 &  0.84 & 24.21 &  0.81 & 21.74 &  0.15 & 20.28 &  0.05 & 19.48 &  0.08 & 18.01 &  0.06 & 17.28 &  0.07 & 16.98 &  0.08 \\
ULAS J005100.87+073514.1  &  12.75361  &   7.58724 & 24.35 &  0.92 & 24.01 &  0.37 & 22.00 &  0.09 & 20.18 &  0.04 & 19.24 &  0.06 & 17.08 &  0.02 & 15.82 &  0.02 & 15.22 &  0.02 \\
ULAS J005235.98+141555.4  &  13.14991  &  14.26539 & 24.46 &  0.85 & 22.41 &  0.13 & 21.00 &  0.06 & 19.86 &  0.03 & 19.22 &  0.05 & 18.01 &  0.05 & 17.23 &  0.06 & 16.96 &  0.07 \\
ULAS J005608.18+052041.9  &  14.03406  &   5.34497 & 23.49 &  0.77 & 23.81 &  0.37 & 22.10 &  0.13 & 20.47 &  0.04 & 19.31 &  0.07 & 17.93 &  0.05 & 17.14 &  0.04 & 16.83 &  0.06 \\
ULAS J005618.98+084641.0  &  14.07907  &   8.77805 & 22.62 &  0.52 & 21.81 &  0.10 & 20.19 &  0.04 & 19.47 &  0.03 & 19.06 &  0.06 & 18.00 &  0.06 & 17.09 &  0.04 & 16.88 &  0.06 \\
ULAS J005859.41+134710.6  &  14.74753  &  13.78627 & 23.58 &  0.70 & 23.31 &  0.30 & 21.69 &  0.08 & 20.18 &  0.03 & 19.32 &  0.06 & 18.09 &  0.06 & 17.35 &  0.06 & 17.03 &  0.08 \\
ULAS J005908.48+085015.2  &  14.78533  &   8.83756 & 23.24 &  0.70 & 21.25 &  0.05 & 19.87 &  0.03 & 19.29 &  0.04 & 18.88 &  0.05 & 18.02 &  0.06 & 17.12 &  0.04 & 16.85 &  0.06 \\
ULAS J010008.14+143019.1  &  15.03390  &  14.50529 & 24.83 &  0.70 & 21.71 &  0.07 & 20.34 &  0.03 & 19.49 &  0.02 & 18.98 &  0.05 & 17.82 &  0.05 & 17.09 &  0.06 & 16.79 &  0.06 \\
ULAS J010133.53+160158.7  &  15.38973  &  16.03297 & 24.49 &  1.08 & 22.36 &  0.23 & 20.24 &  0.07 & 19.54 &  0.06 & 19.25 &  0.12 & 18.14 &  0.07 & 17.35 &  0.06 & 17.01 &  0.07 \\
ULAS J010418.99+001356.7  &  16.07913  &   0.23243 & 22.81 &  0.34 & 21.29 &  0.04 & 19.97 &  0.02 & 19.30 &  0.02 & 18.89 &  0.04 & 17.90 &  0.04 & 17.07 &  0.05 & 16.82 &  0.06 \\
ULAS J011215.23+060439.4  &  18.06347  &   6.07762 & 24.98 &  0.73 & 22.03 &  0.13 & 20.68 &  0.04 & 19.34 &  0.02 & 18.60 &  0.03 & 17.38 &  0.03 & 16.59 &  0.03 & 16.34 &  0.03 \\
ULAS J011531.60+000344.9  &  18.88165  &   0.06249 & 25.57 &  0.45 & 23.09 &  0.15 & 21.71 &  0.06 & 20.25 &  0.03 & 19.42 &  0.06 & 18.00 &  0.05 & 17.20 &  0.06 & 16.84 &  0.07 \\
ULAS J011641.58+065101.6  &  19.17325  &   6.85044 & 23.34 &  0.74 & 24.15 &  0.56 & 22.11 &  0.17 & 20.36 &  0.05 & 19.36 &  0.07 & 18.13 &  0.06 & 17.25 &  0.05 & 17.01 &  0.06 \\
ULAS J011954.61+133849.6  &  19.97753  &  13.64711 & 25.46 &  0.60 & 22.25 &  0.11 & 20.72 &  0.04 & 19.76 &  0.03 & 19.13 &  0.05 & 18.19 &  0.06 & 17.27 &  0.06 & 17.02 &  0.07 \\
ULAS J012226.41+152428.7  &  20.61004  &  15.40798 & 24.61 &  0.91 & 21.04 &  0.04 & 19.78 &  0.03 & 19.12 &  0.02 & 18.67 &  0.04 & 17.60 &  0.04 & 16.78 &  0.02 & 16.55 &  0.04 \\
ULAS J012254.91+061151.2  &  20.72879  &   6.19757 & 24.59 &  0.94 & 21.46 &  0.06 & 20.10 &  0.02 & 19.35 &  0.02 & 18.96 &  0.04 & 17.87 &  0.04 & 17.05 &  0.04 & 16.84 &  0.05 \\
ULAS J012453.59-004120.4  &  21.22329  &  -0.68899 & 22.58 &  0.26 & 21.59 &  0.06 & 20.14 &  0.04 & 19.29 &  0.02 & 18.90 &  0.03 & 17.84 &  0.06 & 17.09 &  0.04 & 16.81 &  0.06 \\
ULAS J012754.13+064807.5  &  21.97554  &   6.80207 & 22.62 &  0.27 & 23.56 &  0.22 & 22.15 &  0.09 & 20.41 &  0.04 & 19.29 &  0.05 & 17.93 &  0.04 & 17.17 &  0.05 & 16.85 &  0.05 
\enddata
 \label{table:candidates}
 \tablecomments{This is an abbreviated list.}
\end{deluxetable*}
\end{center}
\end{turnpage}

\section{Initial Spectroscopic Campaign Results}\label{sec:results}
In this section, we describe our initial spectroscopic campaign, which included observations using five different spectrographs.  The results from our spectroscopic follow-up follows in Section \ref{sec:discussion}.

\subsection{IRTF--Spex} We obtained the bulk of our NIR observations with SpeX at the NASA Infrared Telescope Facility (IRTF).  We observed a total of 13 M giant candidates during 3 nights.  The SpeX observations span 0.8 to 2.5 {\micron}, covering the $JHK$ bandpasses.  We obtained 11 spectra in SXD mode with the 0$^{\prime\prime}$.8 slit, and two spectra in prism mode with the 0$^{\prime\prime}$.5 slit, resulting in spectral resolutions of $R \sim$ 750 and 150, respectively.  The data were reduced using the IDL package \textsc{Spextool} \citep{2004PASP..116..362C}.  This program flat fields and wavelength calibrates each spectrum before performing an optimal extraction.  Telluric correction and flux calibration was performed by comparing A0V stars observed near each science target to spectra of Vega in the \textsc{xtellcor} package.  The results of the IRTF observing runs are given in Table \ref{table:obs}.

\subsection{Gemini Near--Infrared Spectrograph} \label{sec:gemini}
We obtained 14 near--infrared (NIR) spectra of M giant candidates selected with a preliminary targeting algorithm that did not including proper motions or optical color cuts with the Gemini NIR spectrograph \citep[GNIRS;][]{2006SPIE.6269E.138E} at the Gemini Observatory in Mauna Kea, Hawaii.  Only four of the spectra pass our final criteria and are presented in Table \ref{table:obs}.   These early observations were important for testing our quasar and proper motion selection criteria.  Of the 14 observed candidates, eight were photometrically eliminated as QSOs in our final target selection algorithm and were spectroscopically confirmed as QSOS.  The remaining six candidates were all identified as M dwarfs, with two being flagged as moving, while the other four met all of our criteria.  These four were spectroscopically classified as M dwarfs.  As discussed in Section \ref{sec:contam}, this suggests that proper motion selection has an efficiency of $\lesssim 30\%$.

 \begin{figure*} 
  \centering 
\includegraphics[scale=0.4]{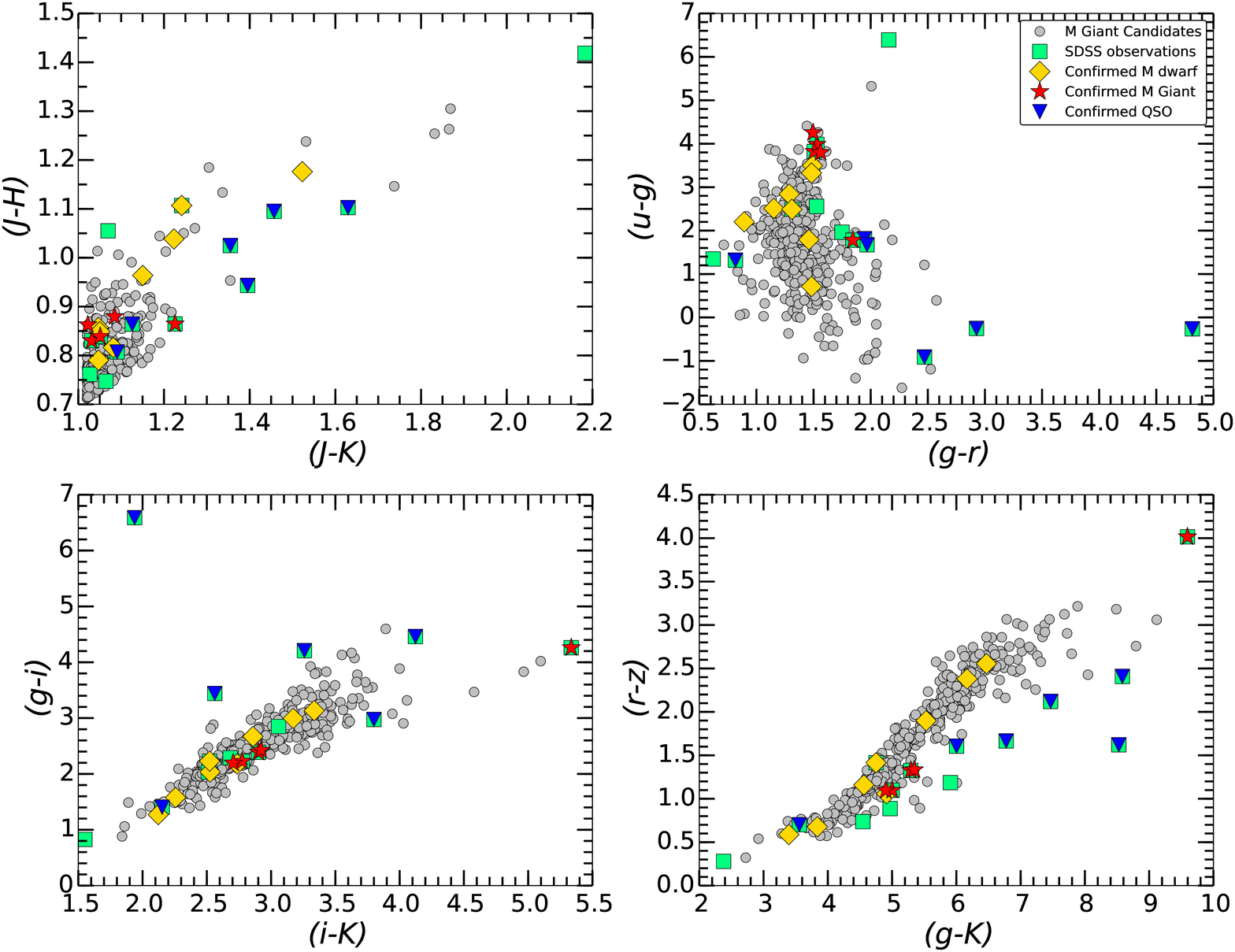}

  \caption{Various color-color diagrams for the M giant candidate sample, along with our spectroscopic observations.  The M giant sample (grey filled squares) has had preliminary spectroscopic followup with SDSS (green filled squares), along with our own observations, yielding QSOs (blue filled triangles), M dwarfs (yellow filled diamonds) and M giants (red filled stars).  It is clear that most QSOs are outliers on the stellar locus, with blue $J-K, g-r$ and $g-K$ colors.  In particular, the $r-z,g-K$ color-color diagram (lower right panel) demonstrates a clear separation from the stellar locus.  The $u-g,g-r$ color-color diagram suggests that M giants are found at redder $u-g$ colors (see also \citealp{2009AJ....137.4377Y}).  There is also suggestive clumping of the confirmed M giants in $r-z,g-K$ color-color space, and this will be further explored in follow-up studies.}
 \label{fig:color-color} 
 \end{figure*}
 
The spectra were cross--dispersed and imaged with the short wavelength camera. The spectra were obtained with the 32 lines per mm grating and the 0.$^{\prime\prime}$30 slit. Each spectrum spans the NIR wavelength regime, from 0.9 to 2.5 {\micron} with a resolution $R \sim 1700$. The spectra were reduced using the \textsc{gnirs} package in the Gemini IRAF\footnote{IRAF is distributed by the National Optical Astronomy Observatory, which is operated by the Association of Universities for Research in Astronomy (AURA) under cooperative agreement with the National Science Foundation.} computing environment. The pipeline corrects for non--linearity, flat fields each image, and performs sky--subtraction from each AB nod. The spectra are wavelength calibrated from Argon lamp spectra, and each order is traced and corrected for curvature before extraction. For telluric correction, we employed the general version of the \textsc{xtellcor} routine first designed for the SpeX spectrograph at the NASA Infrared Telescope Facility \citep{2004PASP..116..362C}. The telluric corrections are computed by comparing the spectrum of an A0V star obtained close to the science target to a spectrum of Vega reddened using the $B-V$ color of the A star. The two spectra are shifted to the same velocity and the telluric correction and flux calibration is computed.

\subsection{SOAR -- Goodman Spectrograph} 
We obtained optical spectra of three M giant candidates with the Goodman High Throughput Spectrograph at the 4.1m Southern Astrophysical Research (SOAR) telescope at Cerro Pach\'{o}n, Chile. These were also targeted with an early version of our selection algorithm.  The targets obtained with SOAR are not included in our final catalog, so this spectroscopic campaign is not included in Table \ref{table:obs}.  The single order spectra were obtained with the 1$^{\prime\prime}$.03 slit and the 930 lines per mm VPH grating centered at 8100 \AA. This setup spans 7250 to 8950 \AA\ with a resolution of $R \sim 4470$. The images were reduced using IRAF routines. Each spectrum was flat-fielded, wavelength calibrated using CuNeAr arc lamp spectra, and extracted. Flux standards were obtained each night for calibration. 

\subsection{FIRE}
We obtained a NIR spectrum of one M giant candidate using the Folded Port Infrared Echellette Spectrograph on the Baade Magellan Telescope \citep[FIRE;][]{2013PASP..125..270S}.  FIRE is a single-object spectrograph capable of two modes:  a single order, low-resolution prism mode, and a cross--dispersed echellette, covering 0.8--2.5 {\micron} ($JHK$) over 21 orders.  On May 15, 2013, we obtained a 1505s exposure of the M giant candidate.  The 0$^{\prime\prime}.6$ slit was used and
aligned with the parallactic angle and the airmass was $1.18$.  An A0V star, HD123233, was obtained for flux calibration and telluric corrections.  The images were reduced using the FIRE reduction software package,
FIREHOSE, which is based on the MASE pipeline \citep{mase}
for the MagE spectrograph \citep{2008SPIE.7014E.169M}. FIREHOSE is used to flat field images, find and trace orders, compute wavelength calibrations from OH telluric emission lines, and extract and combine each spectrum.  A modified version of \textsc{xtellcor} was used to correct for telluric absorption.  Our FIRE observation is listed in Table \ref{table:obs}.

\subsection{SDSS}
In addition to the spectra obtained during our initial observational campaign, 15 M giant candidates had existing SDSS spectra in DR10 \citep{2013arXiv1307.7735A}.  SDSS provides $R \sim 1,800$ resolution optical spectra, covering 3800 to 9200 \AA. Each SDSS spectrum contains a variety of ancillary data, including redshift (or radial velocity for stars), spectral type, and targeting information.  While the pipeline spectroscopic identification is usually reliable, we incorporated detailed visual inspection of all M star spectra.  Of the 15 SDSS spectra, six were classified as QSOs, two were classified as M dwarfs, two as carbon stars, one as an F dwarf, and four M giants by the standard SDSS pipeline \citep{2002AJ....123..485S}.  Color--color diagrams for the objects with SDSS spectra compared to the rest of our candidates are shown in Figure \ref{fig:color-color}.  The F dwarf is much bluer in $g-i, i-K$ than the majority of the sample.  The QSOs are also outliers when compared to the locus of candidates.  Carbon stars and M dwarfs are 
more difficult to distinguish in the $ugrizJHK$ color space, but the $u-g,g-r$ color-color diagram may be useful in discriminating between M giants and other red stars, as previously shown by \cite{2009ApJ...700.1282Y,2009AJ....137.4377Y}.  The results of the SDSS spectroscopy are listed in Table \ref{table:obs}.

\begin{turnpage}
\begin{center}
\begin{deluxetable*}{lrrllrrl}
\tablewidth{0pt}
\tabletypesize{\scriptsize}
 \tablecaption{M Giant Candidates with Spectroscopic Followup}
 \tablehead{
 \colhead{Name} &
 \colhead{RA} &
 \colhead{Dec} &
 \colhead{Instrument} &
\colhead{Sp. Type} &
 \colhead{$RV$ (km s$^{-1}$)} &
 \colhead{$d$ (kpc)} &
 \colhead{Notes} 
}
 \startdata
ULAS J074048.14+261900.2  & 115.20059  &  26.31671 & Spex   & M giant       & $45.0\pm20.0$ &    61 &  \\
ULAS J075554.26+273130.9  & 118.97609  &  27.52525 & SDSS, Spex  & M giant       & $15.9\pm10.2$ &    52 &  \\
ULAS J132441.60-004452.4  & 201.17332  &  -0.74790 & SDSS   & M giant       &$ -8.6\pm 9.7$ &    22 &  \\
ULAS J145254.25-004826.0  & 223.22605  &  -0.80723 & SDSS   & M giant       & $39.3\pm 2.5$ &    28 &  \\
ULAS J151430.44+093722.5  & 228.62683  &   9.62293 & SDSS   & M giant       & $24.1\pm 8.4$ &    92 &  \\
\hline
ULAS J013510.10+004328.8   &  23.79209  &   0.72466 & SDSS   & QSO      &   $z = 1.38$ & \nodata &  \\
ULAS J020549.70+010856.5  &  31.45709  &   1.14904 & SDSS   & F star & $-45 \pm 12$ & 59 &F star with anomalous red colors\\
ULAS J021121.56-003808.5   &  32.83984  &  -0.63568 & GNIRS & M dwarf       &       \nodata &    68 &  \\
ULAS J032746.74-000708.8  &  51.94477  &  -0.11911 & GNIRS  & M dwarf        &       \nodata &   122 &  \\
ULAS J033217.09+002204.0  &  53.07120  &   0.36778 & SDSS      & QSO      &   $z = 4.83$ &   312 &  \\
ULAS J073229.88+250554.0  & 113.12449  &  25.09833 & GNIRS  & M dwarf        &       \nodata &    75 & \\
ULAS J073248.36+272115.5   & 113.20151  &  27.35430 & GNIRS  & M dwarf        &       \nodata &   322 &  \\
ULAS J074030.06+275224.0   & 115.12525  &  27.87333 & Spex   & M dwarf        &       \nodata &    56 &  SXD mode\\
ULAS J083058.67+013448.7  & 127.74446  &   1.58019 & Spex   & \nodata    &       \nodata &    82 & SXD, poor S/N  \\
ULAS J083510.60+022253.4  & 128.79419  &   2.38150 & Spex   & \nodata     &       \nodata &    92 &SXD, inconclusive   \\
ULAS J095021.27+022653.2  & 147.58862  &   2.44810 & Spex   & \nodata     &       \nodata &   112 &SXD, inconclusive  \\
ULAS J100821.40+121348.6  & 152.08916  &  12.23016 & Spex   & \nodata  &       \nodata &    97 & Prism, inconclusive  \\
ULAS J102403.07+105121.6  & 156.01278  &  10.85600 & Spex   & \nodata      &       \nodata &   167 &SXD, inconclusive  \\
ULAS J111523.25+082918.5 & 168.84687  &   8.48848 & SDSS   & QSO      &   $z = 4.77$ &   265 &  \\
ULAS J113931.61+050231.0  & 174.88172  &   5.04195 & SDSS   & \nodata      &  $-43 \pm 16$  &   147 & likely an M dwarf  \\
ULAS J130619.39+023659.1   & 196.58080  &   2.61640 & SDSS   & QSO      &   $z = 4.80$ &  1019 &  \\
ULAS J132856.67+085713.8   & 202.23615  &   8.95383 & SDSS   & QSO      &   $z = 4.14$ &   637 &  \\
ULAS J135418.66-011430.7   & 208.57775  &  -1.24187 & Spex   & \nodata      &       \nodata &    94 &  SXD, inconclusive \\
ULAS J141939.74+101803.3   & 214.91560  &  10.30092 & Spex   & \nodata      &       \nodata &   171 &   SXD, inconclusive\\
ULAS J142526.11+082718.7   & 216.35877  &   8.45518 & SDSS   & QSO      &   $z = 4.95$ &  1111 &  \\
ULAS J142603.74+090326.4  & 216.51559  &   9.05735 & SpeX   & \nodata      &       \nodata &    96 &   SXD, inconclusive\\
ULAS J144239.85+094124.3 & 220.66606  &   9.69008 & Spex   & M dwarf        &       \nodata &    82 & SXD mode \\
ULAS J144631.08-005500.3  & 221.62949  &  -0.91674 & SDSS   & Carbon star &  $79 \pm 4 $  & \nodata & Estimated distance $>$ 5000 kpc \\
ULAS J145112.50+020744.3  & 222.80206  &   2.12897 & Spex, FIRE & M dwarf        &       \nodata &   546 & Observed with prism and SXD mode \\
ULAS J150738.22+104610.6   & 226.90924  &  10.76960 & SDSS     &       Carbon star & $113 \pm 6$  &   151 &  \\
ULAS J221714.27+003346.5  & 334.30948  &   0.56292 & SDSS   & M dwarf        &   $59 \pm 24$ &   331 &

\enddata
 \label{table:obs}
\end{deluxetable*}
\end{center}
\end{turnpage}

\section{Analysis}\label{sec:discussion}

\subsection{Spectroscopic Results}
\subsubsection{Preliminary Classification}
We obtained spectra of 31 M giant candidates using our current algorithm, along with Gemini and SOAR spectra of 13 targets chosen with an earlier version of the pipeline.  Each spectrum was manually inspected.  We did not attempt to classify spectra with low S/N, removing nine stars from our sample.  QSOs were easily identified by their non-stellar SEDs and removed from the sample.  For the SDSS spectra, the pipeline produces a spectral type estimate by comparison to template spectra.  We relied on these spectral types for the identification of non-M stars.  This included one F star with red photometric colors and two carbon stars.  Thirteen M star spectra remained after these cuts, and our dwarf/giant discrimination routine is discussed below.

\subsubsection{M Giant / Dwarf Separation}
\begin{figure*} 
  \centering 
\includegraphics[scale=0.4]{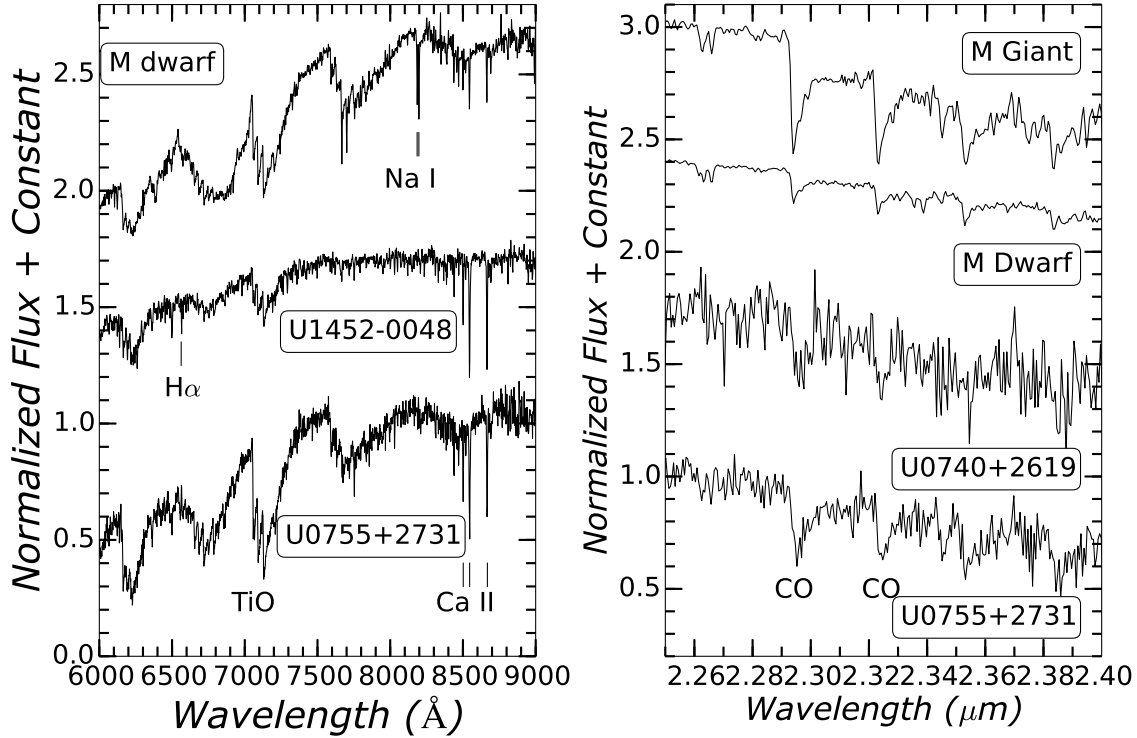}
  
\caption{\textbf{Left Panel:} Optical spectra of two of our M giant candidates, along with an M3 SDSS spectral template \citep{2007AJ....133..531B}.  The main differences between the giants and dwarfs are seen in the differ morphology near the TiO bandheads and the lack of strong Na I absorption at $\sim 8200$ \AA\ in the M giants. \textbf{Right Panel:} $K$ band spectra of two M giants from our study, as well as an M3 giant and M3 dwarf from the IRTF spectral library \citep{2005ApJ...623.1115C}.  Giants possess significantly stronger CO bands than their dwarf counterparts.}
  \label{fig:spectra} 
 \end{figure*}

For the NIR spectra obtained with Gemini, FIRE and Spex, we compared the spectra to M star spectra available in the IRTF spectral library \citep{2005ApJ...623.1115C,2009ApJS..185..289R}.  
The IRTF library contains 86 dwarf and giant spectra, covering 0.8-5.0 {\micron} with a resolution of $R \sim 2,000$.  Each science target and template were normalized to a common scale, and the SEDs were compared.  We examined both the overall agreement in $JHK$ between the science target and template, and the detailed comparison centered on luminosity--sensitive features in the NIR.  Specifically, the Na doublet near 0.82 {\micron}, the Ca triplet near
0.86{\micron}, the Wing-Ford FeH band at 0.99 {\micron}, neutral Na at 1.14 {\micron}, and the CO bandheads in $H$ and $K$ were all examined as they are reliable dwarf$/$giant discriminators \citep{2005ApJ...623.1115C}.  In Figure \ref{fig:spectra}, we compare dwarf and giant spectra from our study.  The luminosity sensitivity of the Na I lines and CO bands are readily apparent (see also Figure 21 of \citealp{2005ApJ...623.1115C}).  Usually, a given science spectrum would have spectral features similar to many dwarf or giant spectra, ensuring a consistent luminosity class identification.  For the identified M dwarfs, the spectral subclass was usually known to $\pm$ 1 spectral subtype.

Optical identification of giants was achieved by comparing the optical spectra from SDSS and SOAR against the \cite{2007AJ....133..531B} M dwarf templates.  These templates are high signal--to--noise averages of M dwarfs with SDSS spectroscopy.  Since the Na I line at 8200 \AA\ is very strong in dwarfs (and weak in giants), the presence (or lack) of significant absorption was used as luminosity indicator \citep{2012AJ....143..114S}.

\subsubsection{Contamination}\label{sec:contam}
While our photometric and proper motion cuts were designed to remove contaminants from the M giant candidates sample, low--mass dwarfs and QSOs, along with other less common contaminants crept into our spectroscopic targets.  It is useful to compare the expected and actual purity of our final sample.  Of the 992 M giants candidates that had clean SDSS and UKIDSS photometry, we estimated $\sim 25\%$ were actually low--mass dwarfs using the $c_1$ color distribution.  Cutting against QSOs removed nearly half the sample, at a fidelity close to $100\%$, as evidenced by the GNIRS spectra.  This effect actually increases the contamination due to M dwarfs in the final sample.  The remaining candidates were expected to be comprised of nearly equal numbers of M dwarfs and M giants (50\% contamination from M dwarfs), plus some unknown fraction of high--redshift ($z > 4$) QSOs and other contaminants.  The proper motion selection removed $\sim 30\%$ of the M dwarfs, but was not nearly as effective at the QSO cut.  
Our spectra followup of 22 M giant candidates with high S/N yielded five bona-fide M giants, along with eight M dwarfs and nine other contaminants.  This suggests that the remaining 404 candidates have a purity of $\sim 20\%$, with contamination being due to both M dwarfs and more exotic contaminants (carbon stars, $z > 4$ QSOs), in roughly equal numbers.  We draw two conclusions from this initial survey.  First, M dwarf contamination is significant within our sample, with M dwarfs outnumbering M giants by at least a factor of two (i.e., 66\% contamination).  This is somewhat larger than our expected fraction of $\sim 50\%$.  Second, our initial study suggests that contamination from exotic contaminants is relatively important in this sample, but we note that these contaminants were exclusively observed by SDSS, which targetted QSOs explicitly.  We expect that as our survey continues, the fraction of M dwarfs will increase relative to the number of exotic contaminants.

\subsubsection{Radial Velocities}
Two of our spectroscopically confirmed M giants have radial velocities (RVs) measured with the SDSS spectroscopic pipeline \citep{2000AJ....120.1579Y, 2002AJ....123..485S}.  The radial velocity of ULAS ~J074048.14+261900.2, which was observed only with Spex, was computed by cross--correlating it against Spex observations of ULAS~J075554.26+273130.9, which was observed with Spex and SDSS, and assuming the SDSS-measured velocity for the template \citep{1979AJ.....84.1511T}.  Due to the lower resolution of the Spex observations, the uncertainty on the RV for the third star is larger than the uncertainty for the SDSS RVs ($\sim 10$ km s$^{-1}$).  

Radial velocities are reported in Table \ref{table:obs}.  We compared the positions, distances and velocities of these M giants to the model predictions of the Sgr stream \citep{2010ApJ...714..229L}.  For each M giant, we computed the median velocity and distance predicted by the model for each wrap of the leading and trailing arms.  We selected 25 of closest spatial matches for each wrap to compute the median model predictions.  These predictions, along with our estimated distances, velocities and arm associations are shown in Table \ref{table:sgr}.

\begin{center}
\begin{deluxetable*}{lrrrrrrl}
\tablewidth{0pt}
\tabletypesize{\scriptsize}
 \tablecaption{Confirmed M Giants in SGR}
 \tablehead{
 \colhead{Name} &
 \colhead{RA} &
 \colhead{Dec} &
 \colhead{$V_{gsr}$} &
 \colhead{$V_{gsr,m}$} &
 \colhead{$d$} &
 \colhead{$d_{m}$} &
 \colhead{Arm} 
 }
\startdata
ULAS J074048.14+261900.2 & 115.20059  & 26.31671 & 227  & 146 & 61 & 39 & $1^{st}$ Trailing \\
ULAS J075554.26+273130.9 & 118.97609  & 27.52525 & 190 &   151 &  52 &  39 & $1^{st}$ Trailing\\
ULAS J132441.60-004452.4 & 201.17332  & -0.74790 & -99 &    -62  &  22 &  20  & $1^{st}$ Trailing\\
ULAS J145254.25-004826.0 & 223.22605 &  -0.80723 & -122 & -109   &  28 &  23 & $1^{st}$ Trailing\\
ULAS J151430.44+093722.5 & 228.62683 & 9.62292 & -148 &  -164  &  92 &  55    & $2^{nd}$ Trailing

\enddata
 \label{table:sgr}
\end{deluxetable*}
\end{center}

\subsection{Photometric Results}

\subsubsection{Galactic Distribution}
\begin{figure*}[htbp] 
  \centering 

\includegraphics[scale=0.9]{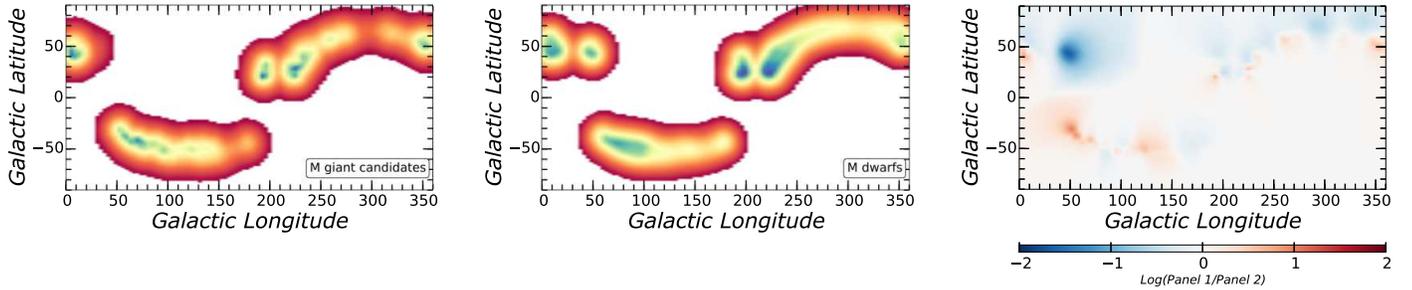}
 
 \caption{\textbf{Left Panel}: The density of M giant candidates on the sky (stars / sq. deg.) estimated using a $k=12$ nearest neighbors algorithm from the \textsc{astroml} python package \citep{2012cidu.conf...47V}. \textbf{Center Panel}: Same as above, but the median density of M dwarfs (from 10,000 random draws matching the M giant candidates in number).  \textbf{Right Panel}: The logarithm of the ratio of the two panels. Note that most of the M giant excesses (shown as red) are far from the Galactic plane, suggesting these may be significant oven-densities.}
  \label{fig:gal_distribution} 
 \end{figure*}

Figure \ref{fig:gal_distribution} shows the distribution of our M giant candidates in Galactic latitude and longitude.  We compared the spatial distributions of the M giant candidates to those of two different samples: 1) a UKIDSS-selected M dwarf sample (point sources with bluer colors and S/N $= $15, see Figure \ref{fig:nir_color}) and 2) a sample composed of 50\% M dwarfs and 50\% QSOs, similar to our contaminants.  If (a fraction of) the M giant candidates are true M giants then the spatial distribution of M giant candidates should be different than that of either comparison sample, with M dwarfs populating the Galactic thin and thick disks, and M giants distributed within the Galactic halo.  

First, we generated each test sample to match the total number of M giant candidates.  Next, we calculated a two--dimensional Kolmogorov--Smirnov test comparing the M giant candidates' Galactic distribution to the contaminant distribution \citep{1987MNRAS.225..155F}.  This test was repeated 10,000 times with a random sub-sample of contaminants for each test, recording the probability each time. For the contaminant sample composed only of M dwarfs, the median probability was 5 $\times 10^{-4}$, strongly suggesting that the M giant candidates and M dwarf samples were not drawn from the same 2D spatial distribution.  We examined the cumulative distribution function for the 10,000 2D KS-test probabilities and found that 99.6\% of all tests had probabilities of $< 5\%$, further strengthening the suggestion that the M giants candidates and M dwarfs are not drawn from the same parent population.  The contaminant sample composed of QSOs and M dwarfs had a similar probability, with 98.4\% of all tests have a probability of $< 5\%$.  Since there is a lack of M giant candidates near ($\ell, b$) $\sim$ (50$^{\circ}$,50$^{\circ}$), we ran the test again, since this may artificially increase the difference between the M giant and contaminant distributions.  We removed stars in this region and recomputed the above test.  No difference between the cleaned samples were detected. 

In Figure \ref{fig:gal_distribution}, we compare the average spatial density (in stars/deg$^2$) of the M giant candidates (left panel), the average map of M dwarf contaminants, computed after 10,000 iterations (center panel) and the logarithm of the ratio of the two distributions (right panel).  The red areas indicate an excess of M giant candidates compared to the average M dwarf spatial density, whereas blue indicates an excess of dwarf stars.  Most of the M giant excess are near an absolute Galactic latitude of $\sim 50$, indicating that many of these stars are likely not distributed within the Milky Way's disk.

\subsubsection{Distances}
\begin{figure*} 
  \centering 
\includegraphics[scale=0.5]{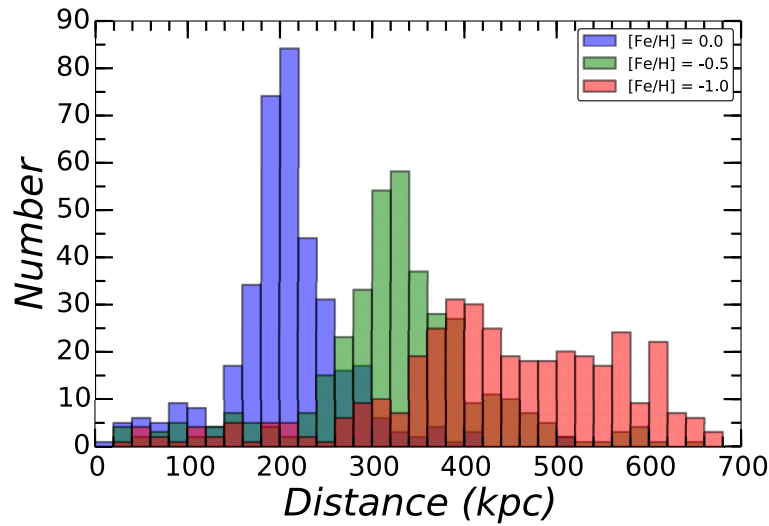}
  \caption{The histogram of distances for M giant candidates in our sample.  Distances were derived using the methods described in \cite{2012ApJ...759..131B}.  The majority of the candidates have distances of 150-300 kpc, with very few stars being found at larger distances.  The confirmed QSOs in our sample have been removed to make this figure, as their computed distances were $> 1000$ kpc.}  
  \label{fig:dist_comparison} 
 \end{figure*}

Photometric parallax relations for M giants are plagued with uncertainty.  Some recent studies have advocated using a single absolute magnitude \citep[i.e., $M_r \sim -2.3$]{2009ApJ...700.1282Y,2012AJ....143..128P}, while other studies advocate simple linear relations \citep{2010ApJ...722..750S}, with large $(\gtrsim 1$ mag) scatter in absolute magnitude \citep[as compared to $\sim$ 0.2 mag for M dwarfs,][]{2010AJ....139.2679B}.    The relations and assumed metallicities are described in Table \ref{table:phot_pi} and were used to estimate distances to all of the M giant candidates in our sample. We adopted the \cite{2012ApJ...759..131B} distance estimates as our fiducial distances.  Briefly, the \cite{2012ApJ...759..131B} distances computed by generating a probability distribution in $M_H, (J-K)$.  This probability distribution is computed using the Padova isochrones, an assumed metallicity, an initial mass function \citep[IMF;][]{2003PASP..115..763C}, and an assumed star formation history (SFH; either constant or exponentially declining).  The number of stars as a function of $M_H, J-K$ is computed, and the probability of a star with $M_H$ is computed for a given $J-K$.  We used the peak of each $M_H, J-K$ probability function to compute our distances.  We refer the reader to Appendix A of \cite{2012ApJ...759..131B} for a further description of this method.  The benefits of this distance estimator is that it does not rely on a single color--magnitude relation, such as the \cite{2000ApJ...542..804N} or \cite{2009ApJ...700.1282Y}  which assumed the color--magnitude relation of the LMC and M71, respectively.  The histogram of distances for M giant candidates in our sample is shown in Figure \ref{fig:dist_comparison}.

To quantify the dependence of distance on the assumed parameters (metallicity and star formation rate), we varied these parameters and re-calculated the distance to each M giant candidate.  We found similar results whether we assumed a constant or exponentially declining SFH.  However, the assumed metallicity of the stellar population has a significant influence on the computed M giant distances, as shown as the different histograms in Figure \ref{fig:dist_comparison}.   For the stars in our sample, the average distance migrates $\sim 100$ kpc outward for each step of $-0.5$ in metallicity.  Given that only the oldest M giants can have low metallicities ($[Fe/H] \sim -1$), we expect that the majority of M giants in our sample are within 200-400 kpc.  The structure seen in Figure \ref{fig:dist_comparison} is due to two major factors:  the distribution of stars in the halo and contamination from nearby M dwarfs.  Given that the halo stellar density distribution decreases as $\sim d^{-3}$ \citep{2008ApJ...673..864J}, while the volume increases by the same factor, we would naively expect a flat or falling distribution of distances.  Thus, the peaks seen in Figure \ref{fig:dist_comparison} are not inherent properties of the halo, and are likely the result of contamination.

\begin{center}
\begin{deluxetable*}{lcll}
\tablewidth{0pt}
\tabletypesize{\scriptsize}
 \tablecaption{Photometric Parallax Relations for M giants}
 \tablehead{
 \colhead{Paper} &
 \colhead{Relation} &
  \colhead{Assumed $[Fe/H]$} &
 \colhead{Source} 
}
 \startdata
\cite{2000ApJ...542..804N}  & $M_K = -5.5$   & $\sim -0.7$ & 2MASS photometry of the LMC \\
\cite{2009ApJ...700.1282Y}  & $M_g = -1.0, M_r \sim -2.3$ & $\sim -0.8$  &  $ugriz$ photometry of M71 \citep{2008ApJS..179..326A,2008AJ....135..682C}  \\
\cite{2010ApJ...722..750S}  & $M_K = 3.26 - 9.42\times(J-K)$   & $ \gtrsim -1$  & Isochrones and ages from $6-13$ Gyr\\
\cite{2012AJ....143..128P}  & $M_r = 0.8, M_i = -1.5, M_z = -3.5$   & $\sim 0$  & \cite{1998PASP..110..863P} spectral templates \\
\cite{2012ApJ...759..131B}  & $P(M_H | J-K)$   & Varies & Assumed SFH, IMF and isochrones
\enddata
 \label{table:phot_pi}
\end{deluxetable*}
\end{center}

\section{Discussion and Conclusions}\label{sec:conclusions}

Using a combination of NIR and optical photometry and spectroscopy from UKIDSS and SDSS, we have assembled a catalog of 404 M giant candidates.  We have identified five M giants with distances from $\sim 20 - 90$ kpc, and are currently targeting fainter M giant candidates.  These fainter stars, if truly M giants, lie at typical distances of $\sim$ 200 kpc (assuming $[Fe/H]$ = 0.0), more distant than any known Milky Way star.  The photometric distance estimates for M giants are highly dependent on their assumed $[Fe/H]$, and increase by $\sim$50\% for each 0.5 dex of decreasing $[Fe/H]$.

We found that optical-NIR cuts to eliminate QSOs and proper motion cuts to eliminate foreground M dwarfs were essential.  Nearly half of the possible M giant candidates were removed when selecting against things with QSO--type colors.  Prior studies that relied solely on NIR photometry to select M giants are only marginally contaminated by QSOs, but contamination becomes important at fainter ($K > 14$) magnitudes.   Early spectroscopic follow-up with Gemini/GNIRS and SOAR confirmed the importance of including both optical photometry and proper motions as a complement to NIR-selected M giants.  In these early samples, objects with NIR$+$optical colors consistent with QSOs and with proper motions consistent with M dwarfs were spectroscopically confirmed to be such with an efficiency of 100\%.

Our initial campaign to follow-up the 404 M giant candidates resulted in 22 candidates with sufficiently high quality spectra to make a robust classification.  5 of these 22 are M giants, which naively implies that $\sim$20\% of our M giant candidates are true M giants.  However, 15 of these 22 spectra are public SDSS spectra targeted because of their unusual colors, for example as indicative of QSOs (6 of the 15) or carbon stars (2 of the 15).  The color distribution of these 15 SDSS targets is significantly different than the overall color distribution of our M giant candidates (Figure \ref{fig:color-color}), so our M giant selection efficiency may be significantly different from 20\%.

In the context of $\Lambda$-CDM cosmology, accretion plays an important role in the assembly and evolution of the Galactic halo.  The archetypal accretion event contributing M giants to the Milky Way's halo is the Sagittarius Dwarf Galaxy \citep{2003ApJ...599.1082M}.  The Sgr dwarf contributed a significant fraction of the halo's M giants, with nearly 75\% of halo M giants (at distances $<$ 50 kpc) coming from the Sgr dwarf \citep{2003ApJ...599.1082M}.  We can use our catalog to loosely constrain the number of Sgr-like accretions that could have built up the Galactic halo by comparing our number of M giant candidates with the expected number of M giants that a Sgr-like accretion would contribute. To do this, we estimate the number of M giants in Sgr by dividing the total luminosity of the system \citep[$\sim 1 \times 10^{7-8} L_{\odot}$; ][]{2010ApJ...712..516N} by the typical luminosity of an M giant \citep[$\sim 1 \times 10^3 L_{\odot}$; ][]{2008ApJS..178...89D} and scaled by the fraction of luminosity produced by M giants.  We employed a variety of  metallicities ($Fe/H = 0.0, -0.2, -0.5, -1.0$) and ages (3, 5, 10 Gyr) to compute the fraction of luminosity in M giants for that model.  For a given cutoff in luminosity, this fraction ranged from 1-30\%.  Thus, given the uncertainties in the total luminosity of Sgr and the fraction of light being emitted by M giants, we expect 10-3000 M giants contributed for each Sgr-accretion event.  Assuming an isotropic distribution of M giants on the sky, we expect $\sim 1-125$ M giants within our UKIDSS sample.  If we naively interpret the 80$\%$ contamination rate of our initial spectroscopic sample, then we expect $\sim 80$ bona-fide M giants within our sample, which is consistent with the accretion of at least a single Sgr-like dwarf.  As explained above, we are biased against the bluest M giants.  We quantified this bias by calculating the fraction of Sgr giants missed by adopting our NIR color cuts.  Using spectroscopically confirmed M giants \citep{2004AJ....128..245M}, we recover $\sim 30\%$ of M giants contained within Sgr.  We note that many of the M giants presented in \cite{2004AJ....128..245M} are early-type M0-M2 giants.  While this cut is necessary to avoid crippling M dwarf contamination, it suggests that the predicted number of M giants in our sample ($\sim 80$), and thus the predicted number of Sgr-type accretion events, is a lower limit for the outer halo.  While all of the M giants presented in this work can be associated with Sgr, M giants found at larger distances ($d > 100$ kpc) are less likely to be well--modeled as Sgr members \citep{2010ApJ...714..229L}.  Other stellar tracers, such as blue horizontal branch stars, and RR Lyrae \citep{2012MNRAS.425.2840D, 2010ApJ...708..717S} will also be important to constraining the total number of accreted Sgr-like dwarfs over the Milky Way's history.

Our candidates lie at distances that are comparable to the virial radius of the Milky Way \citep[$\sim$ 200 kpc;][]{2008ApJ...684.1143X}.  While we only report on radial velocities for five confirmed M giants, our future spectroscopic followup will result in precise radial velocities for most of the stars in our sample.  The kinematics of the outer halo are largely unconstrained but may provide insight into the total mass of the Galaxy and the shape of the Galaxy's gravitational potential.  M giants are some of the most luminous stellar tracers known, and understanding their distribution throughout the Galaxy's halo will be crucial for piecing together the assembly history of our Milky Way.  Significant spectroscopic follow-up of our M giant candidate catalog will yield the first clean map of the Milky Way to its outermost reaches.

\acknowledgements
J.J.B. and B.W. gratefully thank the National Science Foundation for supporting this research under grant NSF AST-1151462.  A.A.W acknowledges funding from NSF grants AST-1109273 and AST-1255568.  A.A.W. also acknowledges the support of the Research Corporation for Science Advancement's Cottrell Scholarship. We thank Rob Simcoe for obtaining FIRE spectra for this work.  We would like to thank all of the observing assistants for their time and patience in assembling our spectroscopic campaign.  We also thank the anonymous referee for their insightful and constructive comments that greatly improved the presentation and content of this study.  This work relied on a bevy of online survey datasets.  This work is based in part on data obtained as part of the UKIRT Infrared Deep Sky Survey.  This research has made use of the VizieR catalogue access tool, CDS, Strasbourg, France.  

Funding for the SDSS and SDSS-II has been provided by the Alfred P. Sloan Foundation, the Participating Institutions, the National Science Foundation, the U.S. Department of Energy, the National Aeronautics and Space Administration, the Japanese Monbukagakusho, the Max Planck Society, and the Higher Education Funding Council for England. The SDSS Web Site is http://www.sdss.org/.

The SDSS is managed by the Astrophysical Research Consortium for the Participating Institutions. The Participating Institutions are the American Museum of Natural History, Astrophysical Institute Potsdam, University of Basel, University of Cambridge, Case Western Reserve University, University of Chicago, Drexel University, Fermilab, the Institute for Advanced Study, the Japan Participation Group, Johns Hopkins University, the Joint Institute for Nuclear Astrophysics, the Kavli Institute for Particle Astrophysics and Cosmology, the Korean Scientist Group, the Chinese Academy of Sciences (LAMOST), Los Alamos National Laboratory, the Max-Planck-Institute for Astronomy (MPIA), the Max-Planck-Institute for Astrophysics (MPA), New Mexico State University, Ohio State University, University of Pittsburgh, University of Portsmouth, Princeton University, the United States Naval Observatory, and the University of Washington.

\appendix
\section{Appendix - UKIDSS SQL Query}
Our SQL query for selecting high quality UKIDSS stellar photometry is given below:

\begin{verbatim}
	SELECT ra, dec, l, b, yAperMag3,yAperMag3Err,j_1AperMag3,j_1AperMag3Err,
		   hAperMag3,hAperMag3Err,kAperMag3,kAperMag3Err, 
		   framesetid, sourceid, epoch
	FROM lasSource
	WHERE 
	       (priOrSec=0 OR priOrSec=frameSetID)
	       AND    yClass   = -1 AND yppErrBits = 0
	       AND    j_1Class = -1 AND j_1ppErrBits = 0
	       AND    hClass   = -1 AND hppErrBits = 0
	       AND    (j_2Class=-1 OR j_2Class = -9999) AND j_2ppErrBits <= 0
	       AND    (kClass=-1   OR kClass = -9999) AND kppErrBits <= 0
	       AND    yXi   BETWEEN -1.0 AND +1.0 AND yEta BETWEEN -1.0 AND +1.0
	       AND    j_1Xi BETWEEN -1.0 AND +1.0 AND j_1Eta BETWEEN -1.0 AND +1.0
	       AND    hXi   BETWEEN -1.0 AND +1.0 AND hEta BETWEEN -1.0 AND +1.0
	       AND    ((kXi BETWEEN -1.0 AND +1.0 AND kEta BETWEEN -1.0 AND +1.0) 
		          OR kXi < -0.9e9) 
	 ORDER BY ra
\end{verbatim}


\end{document}